\documentstyle[aps]{revtex}
\begin{document}
\draft
\title{An algebraic construction of the coherent states of the Morse potential
based on SUSY\ QM }
\author{Bal\'{a}zs Moln\'{a}r\thanks{%
E-mail address: mbalazs@physx.u-szeged.hu} and Mih\'{a}ly G. Benedict%
\thanks{%
E-mail address: benedict@physx.u-szeged.hu}}
\address{Department of Theoretical Physics, Attila J\'{o}zsef University,\\
H-6720 Szeged, Tisza Lajos krt. 84-86, Hungary }
\date{\today }
\maketitle

\begin{abstract}
By introducing the shape invariant Lie algebra spanned by the SUSY ladder
operators plus the unity operator, a new basis is presented for the quantum
treatment of the one-dimensional Morse potential. In this discrete, complete
orthonormal set, which we call the pseudo number states, the Morse
Hamiltonian is tridiagonal. By using this basis we construct coherent states
algebraically for the Morse potential in a close analogy with the harmonic
oscillator. We also show that there exists an unitary displacement operator
creating these coherent states from the ground state. We show that our
coherent states form a continuous and overcomplete set of states. They
coincide with a class of states constructed earlier by Nieto and Simmons by
using the coordinate representation.
\end{abstract}

\section{INTRODUCTION}

Coherent states\cite{Glauber} for systems other than the harmonic oscillator
have attracted great attention for several years
\cite{Nieto,Fukui,Klauder,Perelomov,Gilmore,Gerry,Kronome}. There are a number of different
approaches to this problem and the one presented here is based on the
methods of supersymmetric quantum mechanics (SUSY QM). It is well-known that
this method allows the algebraic treatment of the eigenvalue problems of
Hamiltonians associated with shape invariant potentials \cite
{Gendenstein,Dutt,Levai,Bogar}. Since the SUSY description combined with
the concept of shape invariance can be regarded as a  generalization of the
ladder operator method of the harmonic oscillator, one might think that the
SUSY ladder operators will play an important role in the construction of
coherent states for other, non-harmonic potentials, too. Based on this idea
an algebraic construction of coherent states were proposed by Fukui and
Aizawa \cite{Fukui} for the class of shape invariant potentials having an
infinite number of bound energy eigenstates. Their definition, however, does
not work for potentials, where the number of normalizable energy eigenstates
is finite. Among these latter problems a particular attention deserves the
Morse potential, because it plays an important role in applications like
molecular vibrations and laser chemistry.

In this paper we present a new algebraic method by using the SUSY ladder
operators and shape invariance to obtain coherent states for the
one-dimensional Morse potential. Considering the shape invariant Lie group
spanned by the SUSY ladder operators and the identity, we will introduce a
new orthonormal basis set in the state space, called pseudo number states,
having in a certain extent similar properties with respect of the Morse
potential as the stationary states for the harmonic oscillator. In contrast
with the set of energy eigenstates of the Morse Hamiltonian, this basis is a
complete discrete set of normalizable states. However, this basis does not
diagonalize the Hamilton operator, but tridiagonalizes it. By the help of
these pseudo number states we will define our coherent states in a similar
form as it is usual in the case of harmonic oscillator: 
\begin{equation}
\left| \beta \right\rangle =g(\beta )\sum_{n=0}^{\infty }\frac{\beta ^{n}}{%
\sqrt{\left\{ n\right\} !}}|n\rangle ,  \label{e1}
\end{equation}
where $\beta $ is a complex number, $\left| n\right\rangle $ is an element
of the above mentioned basis, $\left\{ n\right\} !$ is a later specified
generalized factorial and $g(\beta )$ is a normalization term. We will show
that these states satisfy the minimal requirements established by Klauder 
(see in Ref. \cite{Klauder}) to be termed as coherent: 
they are continuous functions of the label $\beta $,
and form an (over)complete set in the Hilbert space. 
It will be also shown that an unitary displacement operator exits in a quite
similar form as in the case of the harmonic oscillator, so that the coherent
states are generated by this operator from the ground state as: 
$\left|\beta \right\rangle =D(\beta )|0\rangle$. 
We note that the coordinate
representation wave functions corresponding to our coherent states have been
obtained earlier by Nieto and Simmons \cite{Nieto} in an entirely different
way. 

\section{THE\ LIE ALGEBRA AND THE PSEUDO NUMBER\ STATES\ OF\ THE\ MORSE\ HAMILTONIAN}

In this work we consider the Morse Hamiltonian: 
\begin{equation}
\widehat{H}(s)=\frac{\widehat{P}^{2}}{2m}+V_{0}\left( s+\frac{1}{2}-\exp
(-\gamma \widehat{X})\right) ^{2},
\end{equation}
where $s$, $V_{0}$ and $\gamma $ are real parameters depending on the
shape of the potential, while the position and the momentum operators obey
the commutation relation $\left[ \widehat{X},\widehat{P}\right] =i\hbar $.
Introducing the dimensionless operators $X=\gamma \widehat{X}$ and
$P=\frac{1}{\sqrt{2mV_{0}}}\widehat{P}$,
and choosing the units so that
$\frac{\gamma\hbar }{\sqrt{2mV_{0}}}=1$,
we have $\left[ X,P\right] =i$ and $\widehat{H}(s)=V_{0}H(s)$ with 
\begin{equation}
H(s)=P^{2}+\left( s+\frac{1}{2}-\exp (-X)\right) ^{2}.  \label{e3}
\end{equation}
From now on we consider this latter $H(s)$ as the Hamiltonian. If $s>0$,  then
there exists a normalizable ground state $\left| \Psi _{0}(s)\right\rangle $
with energy $E_{0}(s)$. As it is known from the theory of SUSY QM one can
introduce the SUSY ladder operators $A(s)$, $A^{\dagger }(s)$ so that $A(s)$
annihilates the ground state: 
\begin{equation}
A(s)\left| \Psi _{0}(s)\right\rangle =0,  \label{e4}
\end{equation}
and the Hamiltonian can be factorized as: 
\begin{equation}
H(s)=A^{\dagger }(s)A(s)+E_{0}(s).  \label{e5}
\end{equation}
In the case of Morse potential the ladder operators
 $A(s)$ and $A^{\dagger }(s)$ can be written as \cite{Dutt}: 
\begin{equation}
\begin{array}{l}
A(s)=s-\exp (-X)+iP, \\ 
A^{\dagger }(s)=s-\exp (-X)-iP.
\end{array}
\label{e9}
\end{equation}
Considering the partner Hamilton operator, $H^{p}(s)=A(s)A^{\dagger }(s)+E_{0}(s)$,
one finds that the Morse potential is shape invariant \cite{Gendenstein,Dutt},
which means: 
\begin{equation}
H^{p}(s)=H(f(s))+R(f(s)), \label{e10}
\end{equation}
with $f(s)=s-1$ and $R(s)=2(s+1)$. Due to this shape invariance property one
can determine the energy eigenstates in an algebraic manner by generating
them from the ground state, as well as the eigenvalues in the following way: 
\begin{equation}
\begin{array}{l}
\left| \Psi _{n}(s)\right\rangle \propto A^{\dagger }(s)\cdots A^{\dagger
}(s-n+1)\left| \Psi _{n}(s-n)\right\rangle , \\ 
E_{n}(s)=E_{0}(s)+
\mathrel{\mathop{\stackrel{n}{\sum }}\limits_{k=1}} R(s-k).
\end{array}
\label{e11}
\end{equation}
%
The Morse potential has only a finite number of bound states (the integer part
of $s+1$), which cannot form a complete set of states in the Hilbert space.
Hence the full quantum description of the Morse potential is impossible  by
restricting oneself only to these bound states. One can of course use the
continuous part of the spectrum of $H$, but instead let us  introduce here
the following infinite series of states: 
\begin{equation}
\begin{array}{l}
\left| 0\right\rangle \equiv \left| \Psi _{0}(s)\right\rangle  \\ 
\left| 1\right\rangle \equiv C_{1}^{-1}A^{\dagger }(s)\left| 0\right\rangle 
\\ 
\vdots  \\ 
\left| n\right\rangle \equiv C_{n}^{-1}A^{\dagger }(s+n-1)\left|
n-1\right\rangle  \\ 
\vdots 
\end{array}
\label{e12}
\end{equation}
where $n$ is a positive integer ($n\in {\bf N}^{+}$) and 
$C_{n}=\sqrt{n(2s+n-1)}$ is a normalization coefficient. 
We would like to emphasize here
that the direction of the parameter shift in (\ref{e12}) is opposite to that
of $f(s)$ appearing in Eqs. (\ref{e10}), and (\ref{e11}). As one can easily
check, the  SUSY ladder operators $A(s)$, $A^{\dagger }(s)$ and the identity
operator span a Lie algebra. 
Since for any $n$ $(n\in {\bf Z})$ we have: 
\begin{equation}
A(s+n)=A(s)+nI\hspace{1cm}(n\in {\bf Z}),  \label{e13}
\end{equation}
the Lie algebra is invariant under the shift of the  shape
parameter $s$. 
An easy calculation shows that the 
SUSY ladder operators satisfy the following commutation relations: 
\begin{equation}
\begin{array}{l}
\left[ A(s+m),A(s+n)\right] =0\hspace{3cm}(n,m\in {\bf Z}), \\ 
\left[ A^{\dagger }(s+m),A^{\dagger }(s+n)\right] =0, \\ 
\left[ A(s+m),A^{\dagger }(s+n)\right] =2sI-\left( A(s)+A^{\dagger}(s)\right) .
\end{array}
\label{e14}
\end{equation}
Eqs. (\ref{e13}-\ref{e14}), are valid for any complex $n$, but we shall exploit
this property only for real, integer $n$.
Using these relations and the fact that $A(s)$ annihilates the ground state: 
$A(s)\left| \Psi _{0}(s)\right\rangle =0$, one can verify that the states
defined in (\ref{e12}) are mutually orthogonal: 
\begin{equation}
\left\langle m\right| n\rangle =\delta _{m,n}.  \label{e15}
\end{equation}
We are going to call these states as pseudo number states of the Morse
potential.

To find the wave functions of our pseudo number states let us introduce a
function of the coordinate variable as $y=2\exp (-x)$. By the help of (\ref{e4})
and (\ref{e12}) we find that the wave functions in question obey the
following recursion relation: 
\begin{equation}
\begin{array}{l}
\varphi _{0}(y):=
\left\langle y\right| 0\rangle ={1\over \sqrt{\Gamma (2s)}}y^{s}\exp(-y/2), \\ 
\varphi _{n}(y):=\left\langle y\right| n\rangle =C^{-1}_{n}(y\frac{\partial }{%
\partial y}+(s+n-1)-\frac{y}{2})\varphi _{n-1}(y).
\end{array}
\label{e15.5}
\end{equation}
Using the Rodrigues' formula for the Laguerre polynomials \cite{Szego} one
can verify that the wave functions appropriate to our pseudo number states
are: 
\begin{equation}
\varphi _{n}(y)=\left( \Gamma (2s){n+2s-1 \choose n}%
\right) ^{-\frac{1}{2}}y^{s}\exp (-y/2)L_{n}^{2s-1}(y).  \label{e16}
\end{equation}
Here $L_{n}^{2s-1}(y)$ denote the generalized Laguerre-polynomials, which
obey the following ortho-normalization relation \cite{Szego}: 
\begin{equation}
\int_{0}^{\infty}
L_{n}^{2s-1}(y)L_{m}^{2s-1}(y)\exp (-y)y^{2s-1}dy=\left( \Gamma (2s){n+2s-1 \choose n}%
\right) \delta_{n,m}.  \label{e16.5}
\end{equation}
Due to the completeness of the Laguerre polynomials with respect of the
weight function $\exp (-y)y^{2s-1}$ \cite{Szego}, the wave functions in (\ref
{e16}) form a complete orthonormal set in the function space $L^{2}\left(
(0,\infty ),\frac{dy}{y}\right) $, (the square integrable functions on the
$(0,\infty )$ interval, with respect of the measure $dy/y$ ), and therefore
the set of the pseudo number states is a complete, orthonormal basis in the
Hilbert space. 

Calculating the matrices of the SUSY ladder operators shifted by an
arbitrary integer $k$, one finds the following matrix elements: 
\begin{equation}
\begin{array}{l}
\left\langle m\right| A(s+k)\left| n\right\rangle =\sqrt{n(2s+m)}\delta
_{m+1,n}-(m-k)\delta _{m,n}, \\ 
\left\langle m\right| A^{\dagger }(s+k)\left| n\right\rangle =\sqrt{m(2s+n)}%
\delta _{m,n+1}-(n-k)\delta _{m,n}, \\ 
\left\langle m\right| H(s)\left| n\right\rangle =\left( 2n(n+s-\frac{1}{2}%
)+E_{0}(s)\right) \delta _{m,n}- \\
\hspace{1cm}-(n-1)\sqrt{n(2s+n-1)}\delta _{m+1,n}-(m-1)%
\sqrt{n(2s+n-1)}\delta _{m,n+1}
\end{array}
\label{e18}
\end{equation}
We see that the fundamental SUSY operators have simple matrices in this new
basis and therefore they can be easily applied in calculations. The matrix
of the Hamilton operator is not diagonal, athough it is quite close to that,
it has nonvanishing elements only in, above and below the diagonal, it is
tridiagonal. As we have noted earlier, in the case of the Morse potential
there is no complete set of normalizable states belonging to the Hilbert
space in which the Hamiltonan would be diagonal.

\section{THE\ COHERENT\ STATES\ OF\ THE\ MORSE\ POTENTIAL}

Let us use now the SUSY operator analogy with the harmonic oscillator, and
define the coherent states of the Morse potential as 
\begin{equation}
\left| \beta \right\rangle =g(\beta )\left\{ I+\sum_{n=1}^{\infty }\frac{%
\beta ^{n}}{n!}A^{\dagger }(s+n-1)\cdots A^{\dagger }(s)\right\} \left|
0\right\rangle ,  \label{e19}
\end{equation}
where $\beta $ is a $c$-number and $g(\beta )$ is a normalization function
to be determined below. Introducing a generalized factorial with the
definition and notation 
\begin{equation}
\left\{ 0\right\} !=1,
\quad \text{ and }\left\{ n\right\} !=%
{\displaystyle {n! \over 2s\cdots (2s+n-1)}}%
=%
{\displaystyle {n+2s-1 \choose n}}^{-1} \quad (n>0), 
\label{e19a}
\end{equation}
the coherent states in (\ref{e19}) can be written by the help of (\ref{e12})
as: 
\begin{equation}
\left| \beta \right\rangle =g(\beta )\sum_{n=0}^{\infty }\frac{\beta ^{n}}{%
\sqrt{\left\{ n\right\} !}}\left| n\right\rangle .  \label{e20}
\end{equation}
To obtain the explicit form of $g(\beta )$ and to find the label space (the
set of allowed $\beta $-s) we set: 
\begin{equation}
1=\langle \beta \left| \beta \right\rangle =g^{2}(\beta )\sum_{n=0}^{\infty }%
\frac{\left| \beta \right| ^{2n}}{\left\{ n\right\} !}=g^{2}(\beta )\left\{
1+\sum_{n=1}^{\infty }\frac{2s\cdots (2s+n-1)}{n!}\left| \beta \right|
^{2n}\right\} .  \label{e20,5}
\end{equation}
The sum in the above expression is convergent if and only if
$\left| \beta \right| <1$, i.e. the label space is the complex open unit disk,
and then the sum in the braces in Eq. (\ref{e20,5}) 
yields $\left( 1-\left| \beta\right| ^{2}\right) ^{-2s}$. 
So we have finally for the coherent states of the Morse potential: 
\begin{equation}
\left| \beta \right\rangle =\left( 1-\left| \beta \right| ^{2}\right)
^{s}\sum_{n=0}^{\infty }\sqrt{%
{\displaystyle {n+2s-1 \choose n}}}
\beta ^{n}\left| n\right\rangle \hspace{1.5cm}
\left( \beta \in {\bf C},\left| \beta \right| <1\right) .  \label{e21}
\end{equation}

The various sets of coherent states that have been introduced in the past
for an arbitrary system have two fundamental common properties established
in Ref. \cite{Klauder}: strong continuity in the label space and
completeness in the sense that there exists a positive measure on the label
space such that the unity operator admits the resolution of unity. Let us
investigate whether our new states introduced in (\ref{e21}) satisfy these
requirements. The first property follows obviously from the definition: if
$\beta \rightarrow \beta ^{\prime }$, where $\beta $, $\beta ^{\prime }$ 
are complex numbers 
$\left| \beta \right| ,\left| \beta ^{\prime }\right| <1$,
then 
$\left\| \left| \beta \right\rangle 
-\left| \beta ^{\prime}\right\rangle \right\| ^{2}\rightarrow 0$. 
To verify the second property, valid for $s>1/2$,
we consider the measure   
$\delta \beta =
{\displaystyle {(2s-1) / \left( 1-\left| \beta \right| ^{2}\right) ^{2}}}
d\mathop{\rm Re}\beta d\mathop{\rm Im} \beta $ ($\left| \beta \right| <1)$ 
and find: 
\begin{equation}
\begin{array}{l}
\mathrel{\mathop{\int }\limits_{\left| \beta \right| <1}}
\left| \beta \right\rangle
\left\langle \beta \right| \delta \beta
%
%
=(2s-1)\stackrel{\infty}{\mathrel{\mathop{\sum }\limits_{n,m=0}}}
{\displaystyle {\left| n\right\rangle \left\langle m\right|
\over \sqrt{\left\{ n\right\} !\left\{ m\right\} !}}}%
\mathrel{\mathop{\int }\limits_{\left| \beta \right| <1}}
(\beta ^{*})^{m}\beta^{n}\left( 1-\left| \beta \right| ^{2}\right) ^{2s-2}d%
\mathop{\rm Re}\beta d\mathop{\rm Im}\beta . 
\label{e22}
\end{array}
\end{equation}
If we introduce polar coordinates in the label space, the integral above can
be calculated easily, and we find that our coherent states form a complete
set and the appropriate form of the resolution of unity is 
\begin{equation}
\mathrel{\mathop{\int }\limits_{\left| \beta \right| <1}}
\left| \beta \right\rangle
\left\langle \beta \right| \delta \beta
 =\pi \sum_{n=0}^{\infty }\left|
n\right\rangle \left\langle n\right| =\pi I.  \label{e23}
\end{equation}
Therefore these states can be regarded as coherent states in the sense of
Ref. \cite{Klauder} too.

We also present here the wave functions corresponding to the coherent states
of the Morse potential. Let us recast Eq. (\ref{e20}) in coordinate
representation in the variable $y=2\exp (-x)$ by using the expression
(\ref {e16}): 
\begin{equation}
\begin{array}{l}
\varphi _{\beta }(y):=\left\langle y\right| \beta \rangle =\left( 1-\left|
\beta \right| ^{2}\right) ^{s}
\stackrel{\infty}{\mathrel{\mathop{\sum }\limits_{n=0}}}
{\displaystyle {\beta ^{n} \over \sqrt{\left\{ n\right\} !}}}%
\langle y\left| n\right\rangle = \\ 
=\left( 1-\left| \beta \right| ^{2}\right) ^{s}
\stackrel{\infty}{\mathrel{\mathop{\sum }\limits_{n=0}}}
{\displaystyle {\beta ^{n} \over \sqrt{\left\{ n\right\} !}}}%
\left( \Gamma (2s)%
{n+2s-1 \choose n}%
\right) ^{-\frac{1}{2}}y^{s}\exp (-y/2)L_{n}^{2s-1}(y)= \\ 
=%
{\displaystyle {\left( 1-\left| \beta \right| ^{2}\right) ^{s} 
\over \sqrt{\Gamma \left( 2s\right) }}}%
y^{s}\exp (-y/2)
\stackrel{\infty}{\mathrel{\mathop{\sum }\limits_{n=0}}}
\beta ^{n}L_{n}^{2s-1}(y).
\end{array}
\label{e24}
\end{equation}
Using the identity for the Laguerre polynomials \cite{Szego}: 
\begin{equation}
\sum_{n=0}^{\infty }w^{n}L_{n}^{\alpha }(y)=(1-w)^{-\alpha -1}
\exp (-{\displaystyle {yw \over 1-w}})
\quad (w\in {\bf C},\left| w\right| <1),
\end{equation}
one obtains that the corresponding wave functions in the $y$ coordinate are: 
\begin{equation}
\varphi _{\beta }(y)=\frac{\left( 1-\left| \beta \right| ^{2}\right) ^{s}}{%
\sqrt{\Gamma \left( 2s\right) }\left( 1-\beta \right) ^{2s}}y^{s}\exp (-%
\frac{y}{2}\frac{1+\beta }{1-\beta }),  \label{e25}
\end{equation}
Here we would like to note that the
wave functions above are essentially the same which have been discovered by
Nieto and Simmons \cite{Nieto} in another way, who called them as
generalized minimal uncertainty coherent states (MUCS) of the Morse
potential. They introduced certain special coordinates in the classical
phase space transforming the trajectories of the bound motions into
ellipses. According to \cite{Nieto}, the MUCS type coherent states are those
which minimize the uncertainty relation of the quantum operators
corresponding to these new classical coordinates called ``natural classical
variables'' in Ref \cite{Nieto}. It is interesting that our algebraic
approach has lead to the same states.
Often the eigenvalue equation that defines the MUCS amounts to the
ladder operator coherent states, if the ground state is a member 
of the minimum uncertainity set. Here we have found that to be the case. 
Otherwise the minimum uncertainity defining equation often yields 
the defining equation for lowering operator squeezed states 
\cite{Truax}.

\section{THE\ DISPLACEMENT\ OPERATOR\ GENERATING 
$\left| \beta \right\rangle$}

In this section we present another interpretation for the coherent states
considered in this paper, by giving the physical meaning of the parameter 
$\beta $. We will show here, that there exists an unitary operator generating
the coherent states from the ground state, thus - according to the
classification of \cite{Nieto} - they can be regarded as displacement
operator coherent states (DOCS), too. Let us consider the wave functions of
the coherent state $\left| \beta \right\rangle $ in the original coordinate
variable $x$. Substituting $y=2\exp (-x)$ in (\ref{e25}), one has: 
\begin{equation}
\varphi _{\beta }(x):=\langle x\left| \beta \right\rangle =\frac{%
e^{-i\varphi }2^{s}}{\Gamma ^{\frac{1}{2}}\left( 2s\right) }e^{-s\left( x-%
\widetilde{x}\right) }\exp \left\{ -e^{-(x-\widetilde{x})}\right\} \exp
\left\{ \frac{-i}{s}\widetilde{p}e^{-(x-\widetilde{x})}\right\} ,
\label{e27}
\end{equation}
where 
$e^{-i\varphi }=\left( {\displaystyle {\left| 1-\beta \right|  \over 1-\beta }}
\right) ^{2s}$ is a phase term and $\widetilde{x}$ and  $\widetilde{p}$ are real
numbers depending on $\beta $: 
\begin{equation}
\begin{array}{l}
\widetilde{x}\equiv \ln (%
\mathop{\rm Re}{\displaystyle {1+\beta  \over 1-\beta }}), \\ 
\widetilde{p}\equiv s
{\displaystyle {\mathop{\rm Im}\frac{1+\beta }{1-\beta }
\over \mathop{\rm Re}\frac{1+\beta }{1-\beta }}}.
\end{array}
\label{e28}
\end{equation}
Calculating the expectation values of the operators $X$ and $P$ in the state 
$\left| \beta \right\rangle $ one obtains: 
\begin{equation}
\begin{array}{l}
\left\langle \beta \right| X\left| \beta \right\rangle =\widetilde{x}%
+\left\langle 0\right| X\left| 0\right\rangle , \\ 
\left\langle \beta \right| P\left| \beta \right\rangle =\widetilde{p}.
\end{array}
\label{e29}
\end{equation}
We can realize that apart from an additive constant the position and
momentum operator expectation values are equal to the numbers $\widetilde{x}$
and $\widetilde{p}$, respectively. (The additive constant is the expectation
value of position in the ground state.) Hence we can introduce a new
labeling for the coherent states by the help of the two numbers $\widetilde{x%
}$ and$\ \widetilde{p}$, as the position and momentum expectation values,
instead of the original complex $\beta $. Therefore this coherent state can
be written as $\left| \widetilde{x},\widetilde{p}\right\rangle $ and the
appropriate label space is ${\bf R}^{2}$ with the measure $d\widetilde{x}d%
\widetilde{p}$ on it. Then the resolution of unity (\ref{e23}) has the
similar form as in the case of the HO:

\begin{equation}
\frac{2s-1}{4 s} \int_{-\infty}^{\infty}\int_{-\infty}^{\infty}
\left|\widetilde{x},\widetilde{p} \right \rangle
\left\langle\widetilde{x},\widetilde{p}\right\rangle
 d\widetilde{x} d\widetilde{p}=\pi I 
\end{equation}
It is
not hard to see that the square of the modulus of the wave function in (\ref
{e27}) is equal to the square of the modulus of the ground state function
shifted along the $x$-axis. Eq. (\ref{e27}) also implies that our coherent
states can be written as 
\begin{equation}
\begin{array}{l}
\left| \widetilde{x},\widetilde{p}\right\rangle =e^{-i\varphi }\exp (-i%
\widetilde{x}P)\exp (-{\displaystyle{i \over s}}%
\widetilde{p}e^{-X})\left| \widetilde{x}=0,\widetilde{p}=0\right\rangle = \\ 
\hspace{1cm}=e^{-i\varphi }\exp (-{\displaystyle{i \over s}}%
\widetilde{p}e^{\widetilde{x}I})\exp (-{\displaystyle{i \over s}}%
\widetilde{p}e^{-X})\exp (-i\widetilde{x}P)
\left| \widetilde{x}=0,\widetilde{p}=0\right\rangle ,
\end{array}
\label{e31}
\end{equation}
where $\left| \widetilde{x}=0,\widetilde{p}=0\right\rangle $ is identical to
the ground state $\left| \beta =0\right\rangle $ being itself a coherent state, as well.
Writing the definitions (\ref{e9}) of the $A(s)$ and $A^{\dagger }(s)$ into
the above expression one obtains: 
\begin{equation}
\begin{array}{l}
\left| \widetilde{x},\widetilde{p}\right\rangle =e^{-i\varphi }
\exp (-i\widetilde{p}I)\exp ({\displaystyle{\widetilde{x} \over 2}}%
\left( A^{\dagger }(s)-A(s)\right) )\exp \left( 
{\displaystyle{i \over 2s}}%
\widetilde{p}\left( A(s)+A^{\dagger }(s)\right) \right) \left|
0\right\rangle , \\ 
\left| \widetilde{x},\widetilde{p}\right\rangle =e^{-i\varphi }
\exp (-{\displaystyle{i \over s}}%
\widetilde{p}e^{\widetilde{x}I})\exp (-i\widetilde{p}I)
\exp \left( {\displaystyle{i \over 2s}}%
\widetilde{p}\left( A(s)+A^{\dagger }(s)\right) \right) \times  \\ 
\hspace{2.4cm}\times \exp ({\displaystyle{\widetilde{x} \over 2}}%
\left( A^{\dagger }(s)-A(s)\right) )\left| 0\right\rangle .
\end{array}
\label{e32}
\end{equation}
Considering the case of the harmonic oscillator it inspires to introduce a
displacement operator in the following way 
\begin{equation}
D(\widetilde{x},\widetilde{p})=e^{-i\varphi }\exp (-i\widetilde{p}I)
\exp ({\displaystyle{\widetilde{x} \over 2}}%
\left( A^{\dagger }(s)-A(s)\right) )
\exp \left( {\displaystyle{i \over 2s}}%
\widetilde{p}\left( A(s)+A^{\dagger }(s)\right) \right) .  \label{e33}
\end{equation}
Then according to (\ref{e33}) the coherent states are created by this 
$D(\widetilde{x},\widetilde{p})$ operator from the ground state: 
\begin{equation}
\left| \beta \right\rangle \equiv \left| \widetilde{x},\widetilde{p}%
\right\rangle =D(\widetilde{x},\widetilde{p})\left| 0\right\rangle .
\label{e34}
\end{equation}
From the definition in (\ref{e34}), which is the obvious generalization of the
case of \ the oscillator, it follows that the $D(\widetilde{x},\widetilde{p})
$ operators are unitary for arbitrary $\widetilde{x}$, $\widetilde{p}$ and
it also proves that our states belong to the DOCS category according 
to \cite{Nieto}.

\section{CONCLUSIONS\ AND\ FINAL\ REMARKS}

By using the set of generalized creation and annihilation operators we have
introduced the complete orthonormal set of pseudo number states for the
Morse potential. Then, with a construction similar to the case of the harmonic
oscillator, we have introduced the set of coherent states depending on the
complex parameter $\beta $. We have shown how this parameter is connected
with the expectation values of the coordinate and the momentum, and have
determined the unitary displacement operator generating our coherent states from
the ground state.  

In our construction of the states $\left| \beta \right\rangle $ the
fundamental role has been played by the shape invariant Lie algebra (\ref
{e14}) spanned by the SUSY ladder operators plus the identity. The pseudo
number states have been are generated by its elements $A^{\dagger }(s+n)$, while 
The unitary
displacement operators, $D(\widetilde{x},\widetilde{p})$, that 
create the coherent states,  are the elements
of the corresponding Lie group obtained by exponentiating of the SUSY ladder algebra. 
As it can be simply shown, this algebra, which is solvable, but not nilpotent, 
is not isomorphic to the Heisenberg-Weyl algebra of the harmonic oscillator. 
Moreover our new SUSY ladder algebra is not isomorphic either to the class of
$so(2,1)\cong su(1,1)\cong sp(1,{\bf R})$, which has been applied to the
various recent treatments of the Morse potential 
\cite{Gerry,Alhassid}.
Therefore the SUSY ladder operator algebra and
the coherent states presented here can be regarded as a new algebraic
viewpoint for the description of the one-dimensional Morse potential. 

We also think that our construction will be useful in describing 
molecular interactions with the electromagnetic field.

\section{ACKNOWLEDGEMENTS}

The authors thank F. Bartha, F. Bog\'ar, L. Feh\'{e}r and G. L\'evai 
for the useful discussions and remarks. This
work was supported by the National Research Foundation of Hungary (OTKA)
under contract No. T22281.

\end{document}